\documentclass{article}

\PassOptionsToPackage{numbers, compress}{natbib}

\usepackage[preprint]{nips_arxiv}

\usepackage[utf8]{inputenc} % allow utf-8 input
\usepackage[T1]{fontenc}    % use 8-bit T1 fonts
\usepackage{hyperref}       % hyperlinks
\usepackage{url}            % simple URL typesetting
\usepackage{booktabs}       % professional-quality tables
\usepackage{amsfonts}       % blackboard math symbols
\usepackage{nicefrac}       % compact symbols for 1/2, etc.
\usepackage{microtype}      % microtypography
\usepackage{xcolor}         % colors
\usepackage{multirow}
\usepackage{graphicx}
\usepackage{wasysym}
\usepackage{floatrow}
\newfloatcommand{capbtabbox}{table}[][\FBwidth]
\usepackage{wrapfig}
\usepackage{enumitem}
\usepackage{CJK}
\usepackage{subcaption}  % 用于子图

\usepackage{lineno}
\usepackage{tabularx}
\usepackage{threeparttable}
\usepackage{authblk}

\title{EmotionTalk: An Interactive Chinese Multimodal Emotion Dataset With Rich Annotations}

\author[1,\dag]{Haoqin Sun}
\author[1,\dag]{Xuechen Wang}
\author[1]{Jinghua Zhao}
\author[1]{Shiwan Zhao}
\author[1]{Jiaming Zhou}
\author[1]{Hui Wang} 
\author[1]{Jiabei He}
\author[1]{Aobo Kong}
\author[2]{Xi Yang}
\author[2]{Yequan Wang}
\author[2]{Yonghua Lin}
\author[1,\thanks{Corresponding author. $\dag$ These authors contributed equally to the manuscript.}]{Yong Qin}

\affil[1]{Nankai University}
\affil[2]{Beijing Academy of Artificial Intelligence}

\affil[ ]{\texttt{sunhaoqin@mail.nankai.edu.cn, qinyong@nankai.edu.cn}}

\begin{document}

\maketitle

% \th[$\dag$]{corresponding author.}
% \footnotetext{These authors contributed equally to the manuscript.}

\begin{abstract}
  In recent years, emotion recognition plays a critical role in applications such as human-computer interaction, mental health monitoring, and sentiment analysis. While datasets for emotion analysis in languages such as English have proliferated, there remains a pressing need for high-quality, comprehensive datasets tailored to the unique linguistic, cultural, and multimodal characteristics of Chinese. In this work, we propose \textbf{EmotionTalk}, an interactive Chinese multimodal emotion dataset with rich annotations. This dataset provides multimodal information from 19 actors participating in dyadic conversational settings, incorporating acoustic, visual, and textual modalities. It includes 23.6 hours of speech (19,250 utterances), annotations for 7 utterance-level emotion categories (happy, surprise, sad, disgust, anger, fear, and neutral), 5-dimensional sentiment labels (negative, weakly negative, neutral, weakly positive, and positive) and 4-dimensional speech captions (speaker, speaking style, emotion and overall). The dataset is well-suited for research on unimodal and multimodal emotion recognition, missing modality challenges, and speech captioning tasks. To our knowledge, it represents the first high-quality and versatile Chinese dialogue multimodal emotion dataset, which is a valuable contribution to research on cross-cultural emotion analysis and recognition. Additionally, we conduct experiments on EmotionTalk to demonstrate the effectiveness and quality of the dataset. It will be open-source and freely available for all academic purposes. The dataset and codes will be made available at: https://github.com/NKU-HLT/EmotionTalk.
\end{abstract}

\section{Introduction}
Multimodal emotion recognition (MMER) has become a key focus in artificial intelligence, integrating speech, vision, and text to capture the complexity of human emotions. It drives advancements in applications like virtual assistants, online education, and mental health monitoring. However, most research relies on English datasets, with Chinese resources remaining scarce. Existing datasets often face issues such as low quality, limited scale, and incomplete modalities, hindering model performance. Therefore, the development of a high-quality Chinese multimodal emotion recognition dataset is of critical importance to advance research in this field.

Traditional emotion recognition tasks include unimodal / multimodal emotion recognition on isolated utterances~\cite{liu22aa_interspeech,liu2023multi,sun2024iterative,sun2024fine} and conversational emotion recognition~\cite{shi2020dimensional,shi2023emotion}. The former relies on a single modality or integrates multimodal information for emotion recognition. For example, MISA~\cite{hazarika2020misa}utilizes modality-invariant and modality-specific representations to fuse multimodal information. FDMER~\cite{yang2022disentangled} extends MISA by incorporating tailored constraints and adversarial learning strategies to effectively capture multimodal information. The latter focuses on analyzing emotional changes by examining context and emotional evolution. DialogueRNN~\cite{majumder2019dialoguernn} extracts emotional information from conversations by modeling the speaker, context, and emotions within the dialogue. DialogueGCN~\cite{ghosal2019dialoguegcn} and MMGCN~\cite{wei2019mmgcn} leverage graph-based networks to model the dependencies within dialogues. As emotion recognition research continues to advance, researchers introduce new tasks such as emotion recognition in missing modality scenarios~\cite{zeng2022mitigating,zeng2022tag,sun2025enhancing} and emotion caption~\cite{xu2024secap,liang2024aligncap}, driven by evolving applications and practical requirements. TATE~\cite{zeng2022tag} utilizes a Tag-Assisted Transformer Encoder network to guide the model in focusing on different missing cases by encoding specific tags for the missing modalities. As for the emotion captioning task, it is first proposed by SECap~\cite{xu2024secap}, which employs the SSL model and LLaMA to generate emotion descriptions. 

However, these studies use different datasets, and while they perform well in their respective experiments, directly comparing their performance remains challenging. This is mainly due to significant differences in dataset scale, annotation methods, modality combinations, and dialogue structures, which affect model applicability and generalization. For instance, popular multimodal benchmarks like IEMOCAP~\cite{busso2008iemocap}, MELD~\cite{poria2019meld}, CMU-MOSEI~\cite{zadeh2018multimodal}, and CH-SIMS~\cite{yu2020ch} have been widely used but are primarily in English, with varying emotion category definitions and annotation standards, limiting cross-lingual and cross-cultural applicability. Specifically, IEMOCAP, one of the most widely used emotion recognition datasets, has only 5,531 usable samples, totaling around 7 hours of data. Meanwhile, the CH-SIMS dataset integrates multiple modalities but lacks clear emotional discrete label definitions and dialogue contexts. Additionally, in emotion captioning, most research relies on unpublished datasets, leading to a lack of a standardized, open benchmark, which hinders reproducibility and broader application.

Due to the vast availability of resources on the internet, obtaining these resources has become relatively easy, and previous Chinese multimodal emotion datasets have largely relied on publicly available data. However, these datasets still have limitations in terms of quality and task coverage, and cannot fully meet the demands of multimodal emotion research. Therefore, constructing a high-quality, comprehensive emotion dataset is particularly important. Such a dataset would not only address the gaps in existing resources but also provide a unified benchmark for emotion-related tasks, contributing to the advancement of research in the field of affective computing. In this paper, we construct an large-scale interactive Chinese multimodal emotion dataset with fine-grained labels and emotional speaking style captions, \textbf{EmotionTalk}, in which the data are contributed by 19 professional actors, ensuring the naturalness and authenticity of the emotion expression. The dataset is in the form of dialogues, containing 23.6 hours of data and 19,250 utterances, along with corresponding labels that support various emotion tasks, including 7 discrete labels, 5 dimensional labels, and 4 caption labels. To the best of our knowledge, EmotionTalk is the first large-scale, comprehensive, recorded interactive Chinese multimodal emotion dataset. We further conduct experiments on unimodal emotion recognition, multimodal emotion recognition, and emotion caption tasks to validate the effectiveness and applicability of the constructed dataset. These experiments not only demonstrate the dataset's performance across different emotion tasks but also highlight its potential to support diverse model development and evaluation.

\section{Related Work}
\subsection{Related Datasets}
% wxc
Table \ref{dataset} presents the datasets which are commonly used in the field of multimodal emotion recognition, all of which consist of video, audio and text modalities.

\textbf{English Datasets:} The CMU-MOSEI~\cite{zadeh2018multimodal} and MELD~\cite{poria2019meld} datasets provide large-scale multimodal data sourced from YouTube and TV shows, covering tasks such as discrete emotion classification and continuous sentiment intensity prediction. These datasets are advantageous due to their rich emotional labeling, but they are primarily derived from entertainment content, where emotional expressions tend to be exaggerated. As such, they may not fully capture the natural emotional expressions encountered in real-life situations. In contrast, the CREMA-D \cite{cao2014crema}, RAVDESS \cite{livingstone2018ryerson}, IEMOCAP~\cite{busso2008iemocap} and MSP-IMPROV~\cite{busso2016msp} datasets are based on actor performances and emotion training, with IEMOCAP and MSP-IMPROV consist of conversational data, whereas CREMA-D and RAVDESS record non-dialogue data. These datasets offer higher-quality emotional data. However, due to their scripted nature, the limitations of the dialogue scripts can affect the actors' performances, leading to emotional expressions that may feel unnatural or overly theatrical.

\newcommand{\datasetfontsize}{\fontsize{8}{9}\selectfont} % 自定义表格字体大小

\begin{table*}
    \datasetfontsize
    \centering
\begin{tabularx}{\textwidth}{lccccccc}
\toprule
\textbf{Dataset} & \textbf{Modality} & \textbf{Dialogue} & \textbf{Sources} & \textbf{Emo-label} & \textbf{Des.} & \textbf{Language} & \textbf{Utts} \\ 
\midrule
CMU-MOSI \cite{zadeh2016mosi} &$a, v, l$ &No& YouTube & 7 Dim. & No & English & 2,199 \\ 
CMU-MOSEI \cite{zadeh2018multimodal} &$a, v, l$ &No& YouTube & 7 Disc. / 5 Dim. & No & English & 22,856 \\ 
MELD \cite{poria2019meld} &$a, v, l$ &Yes& TVs & 7 Disc. & No & English & 13,708 \\ 
CREMA-D \cite{cao2014crema} &$a, v, l$&No& Act & 6 Disc. & No & English & 7,442 \\ 
RAVDESS \cite{livingstone2018ryerson}&$a, v, l$&No & Act & 8 Disc. & No & English & 7,356 \\ 
IEMOCAP \cite{busso2008iemocap}&$a, v, l$&Yes & Act & 5 Disc. & No & English & 7,433 \\ 
MSP-IMPROV \cite{busso2016msp}&$a, v, l$&Yes& Act & 5 Disc. & No & English & 8,438 \\ 
CH-SIMS \cite{yu2020ch}&$a, v, l$&No & Movies, TVs & 5 Dim. & No & Mandarin & 2,281 \\ 
MER-MULTI \cite{lian2024merbench} &$a, v, l$&No & Movies, TVs & 6 Disc. & No & Mandarin & 3,784 \\ 
M$^3$ED \cite{zhao2022m3ed}&$a, v, l$ &Yes& TVs & 7 Disc. & No & Mandarin & 24,449 \\ 
MC-EIU\_ch \cite{liu2024emotion}&$a, v, l$ &Yes& TVs & 7 Disc. & No & Mandarin & 11,003 \\ 
\textbf{EmotionTalk}& \textbf{$a, v, l$} &\textbf{Yes} & \textbf{Act} & \textbf{7 Disc. / 5 Dim.} & \textbf{Yes} & \textbf{Mandarin} & \textbf{19,250} \\ 
\bottomrule
\end{tabularx}
\caption{Summary of multimodal emotion datasets.}
\label{dataset}
\end{table*}

\textbf{Chinese Datasets:} 
Currently, there have been some preliminary research efforts in the field of multimodal emotion datasets based on Mandarin For example, the CH-SIMS~\cite{yu2020ch} and MER-MULTI \cite{lian2024merbench} dataset use five continuous emotion labels and six discrete emotion labels respectively, making it suitable for multimodal sentiment analysis on isolated utterances spoken in Mandarin. However, both of them lack dialogue scenarios, overlooking the emotional changes multi-turn interactions. In contrast, datasets like M$^3$ED~\cite{zhao2022m3ed} and MC-EIU$_{ch}$~\cite{liu2024emotion} have made progress in terms of dialogue-level data, making it possible for supporting multimodal emotion recognition in conversations. Moreover, M$^3$ED and MC-EIU$_{ch}$ have been significant progress regarding the scale of the data.

Overall, the existing Mandarin multimodal emotion datasets still have gaps compared to the English datasets. Most of the Mandarin datasets focus on relatively simple emotion labels or data sourced from the internet, which limits both data quality and emotional annotation. As a result, the mentioned datasets struggle to support some specific emotion tasks, such as missing modality scenarios and emotion captioning.

\subsection{Related Methods}
\subsubsection{Multimodal Emotion Recognition}
Multimodal emotion recognition typically involves identifying the speaker's emotions from an isolated utterance or a dialogue. The methods of multimodal feature fusion play a crucial role in utterance-level emotion recognition. \citet{yang-etal-2023-self} propose a context representation module and a self-adaptive path selection module, obtaining final integrated multimodal features. \citet{fan2024atta} design an attention aggregation network and a auxiliary uni-modal classifier to align shared emotional information across modalities. Design an attention aggregation network and a auxiliary uni-modal classifier to align shared emotional information across modalities. \citet{10109845} propose a transformer-based model with self-distillation for conversational emotion recognition. The framework could capture intra- and inter-modal interactions and obtain more expressive context-sensitive features. \citet{9747397} propose a graph-based dynamic fusion network to reduce redundancy in multimodal interactions and model the context features in a conversation. In addition, since discrete emotion labels are unable to capture the complexity of emotional intensity, some work focuses more on continuous multimodal emotion recognition. For instance, \citet{yang-etal-2024-clgsi} propose a multimodal emotion analysis framework based on contrastive learning guided by sentiment intensity. The effectiveness has been validated on the CMU-MOSEI and CH-SIMS datasets, which include labels related to emotional intensity.

\subsubsection{Emotion Captioning}
To break through the limitations of traditional emotion recognition and capture richer emotional information, the task of emotion captioning has gradually emerged. For example, \citet{xu2024secap} propose a speech emotion captioning framework, effectively discrbing speech emotions with the help of LLaMA \cite{touvron2023llamaopenefficientfoundation} and HuBERT \cite{10.1109/TASLP.2021.3122291}. \citet{liang-etal-2024-aligncap} design a network named AlignCap to align speech emotion captioning to human preferences. This approach improves the generalization on unseen speech, obtaining stronger performance to other zero-shot emotion captioning methods. In addition, captioning-related tasks have also found some applications in TTS. For an instance, \citet{kawamura2024librittspcorpusspeakingstyle} uses speaker captions and speaking style captions, which contain emotional information, to train a TTS system. This TTS system demonstrates higher naturalness and word accuracy with the help of speaking style and speaker identity prompts.

\section{Dataset Description}
\label{sec:dataset}
In this section, we introduce a large-scale, comprehensive, recorded interactive Chinese multimodal emotion dataset, EmotionTalk. We describe data Collection, annotation and statistics in detail.

\subsection{Data Collection}
Compared to other Chinese multimodal datasets, our data is recorded by professional actors from the drama department, ensuring more authentic and natural emotional expressions, thus better simulating spontaneous emotional behavior in real-world environments. At the same time, this process is more challenging and time-consuming.

To ensure data diversity, we create situational scripts that simulate real-life interpersonal interactions in a dialogue format. The scripts are inspired by television plotlines or generated by large language models (LLMs) and then manually reviewed for quality. Each dialogue scene features two characters, with multiple rounds of interaction designed to capture the dynamic changes in emotions over time. The scripts include varying emotional intensities, ranging from lighthearted conversations to intense emotional conflicts, fully showcasing the diversity of emotional expression. 

The scripts cover multiple life themes, such as friendship, family, workplace, and patient-caregiver interactions. The friendship theme includes dialogues about joy, arguments, and reconciliation, highlighting support and conflict. The workplace theme addresses complex emotions like collaboration, competition, pressure, and misunderstandings. Each theme is intricately designed, with the family theme encompassing scenarios like family arguments, holiday reunions, and farewells, reflecting emotions like warmth, anger, and sadness. The language style is tailored to each theme: friendly dialogues are casual and natural, while workplace conversations are more formal and serious, effectively simulating real-world language environments and inspiring authentic emotional performances from the actors, thereby enhancing the script's emotional depth and immersion. It is important to note that actors are not constrained by the script itself. Instead, they are encouraged to express their authentic emotions based on the theme and subject matter. Furthermore, considering that performers cannot sustain a single emotion or continuously portray the same emotion for extended periods, each dialogue lasts approximately 2 minutes.

\subsection{Annotation}
\label{sec:annotation}
To ensure the high quality and diversity of the dataset, we design a rigorous data annotation process, incorporating multi-dimensional annotations for emotion categories, emotion intensity, and emotional speaking style caption. The detailed annotation process is outlined below:

\textbf{Emotion Category:} 
For each sample, we design a multi-step annotation process with cross-validation by \textit{N (N = 5)} annotators. The emotion category annotation is based on the basic emotion theory commonly used in psychological research and covers \textit{K (K = 7)} widely recognized emotion categories: happiness, surprise, sadness, disgust, anger, fear, and neutral. To prevent interference between different modalities and avoid potential confusion, we follow the annotation principles of CH-SIMS, requiring annotators to only view the information from the current modality without performing simultaneous annotations across multiple modalities. The annotation process follows a predefined sequence, starting with text, followed by audio, silent video, and finally multimodal integration. Each emotion annotation consists of a emotion category $y_i$ and a confidence score $c_{i}$, of which is set to 0.1, 0.3, 0.5, 0.7, and 0.9, to quantify the annotator's confidence in their judgment. The formula for calculating the weighted confidence score $x_k$, \textit{k = \{1, …, K\}} for each category of a sample is as follows:
\[
x_k = \frac{1}{N_k} \sum_{i=1}^{N} \mathbb{I}(y_i = k) \cdot c_i
\]
where \(k\) represents the emotion category, \(N_k\) is the number of annotations for category \(k\), \(\mathbb{I}(y_i = k)\) is an indicator function, which equals 1 if the label \(y_i\) assigned by annotator \(i\) is equal to category \(k\), and 0 otherwise.

Thus, the final emotion category $y$ is calculated as follows:
\[
y = \arg \max_{k} x_k
\]
where $argmax$ represents selecting the category \(k\) that corresponds to the maximum weighted confidence \(x_k\) as the final category label.

For annotations with low confidence scores or inconsistent annotations, expert reviewers are consulted to make final determinations, ensuring the accuracy and reliability of the annotations.

\textbf{Emotion Intensity:} 

To more accurately quantify the intensity of emotional expressions, we have designed a multimodal-based emotion intensity annotation process aimed at quantitatively labeling the emotional polarity (positive, negative, neutral) and its intensity in utterances. For each audio clip, five annotators will be assigned, and each annotator will evaluate the emotional state as -2 (strongly negative), -1 (weakly negative), 0 (neutral), 1 (weakly positive), or 2 (strongly positive). The annotation results from the five annotators are then averaged to obtain a continuous label that contains emotion intensity information. The final labeling results will be one of the following values: \{-2.0, -1.8, -1.6, -1.4, -1.2, -1.0, -0.8, -0.6, -0.4, -0.2, 0.0, 0.2, 0.4, 0.6, 0.8, 1.0, 1.2, 1.4, 1.6, 1.8, 2.0\}. Smaller values indicate higher negativity, while larger values indicate higher positivity.

\textbf{Emotional Speaking Style Caption:} 
The most crucial aspect is that we design an innovative speech annotation process aimed at comprehensively describing the emotional and expressive features in speech. This annotation system covers four distinct dimensions: speaker caption, speaking style caption, emotion caption, and emotional speaking style caption. In this process, we first annotate the speaker's voice quality, focusing on features such as warmth, hoarseness, and clarity. Next, the speaking style caption emphasizes annotating speech rate, intonation, stress, and pauses, highlighting the individual's linguistic habits in communication. Emotion caption details the type and intensity of emotions conveyed in the speech. Finally, the integrated caption combines the features from the speaker, speaking style, and emotion dimensions to generate a comprehensive and accurate semantic summary. To further enhance the diversity and expressive capability of the annotations, we employ a LLM to expand the integrated descriptions, generating five semantically consistent but stylistically varied versions, thus ensuring data diversity and flexibility.

\subsection{Statistics}
The dataset is composed of 744 dialogues, with 19,250 utterances for each unimodal modality: text, audio, and video. The audio data have a total duration of 23.6 hours, with an average length of 4.4 seconds per utterance. The textual data comprises 469,387 characters in total, averaging approximately 24 characters per sentence. In the following parts, we provide a detailed description of each modality.

\begin{figure}[t]
    \centering
    \begin{subfigure}{0.37\linewidth}
        \centering
        \includegraphics[width=1\linewidth]{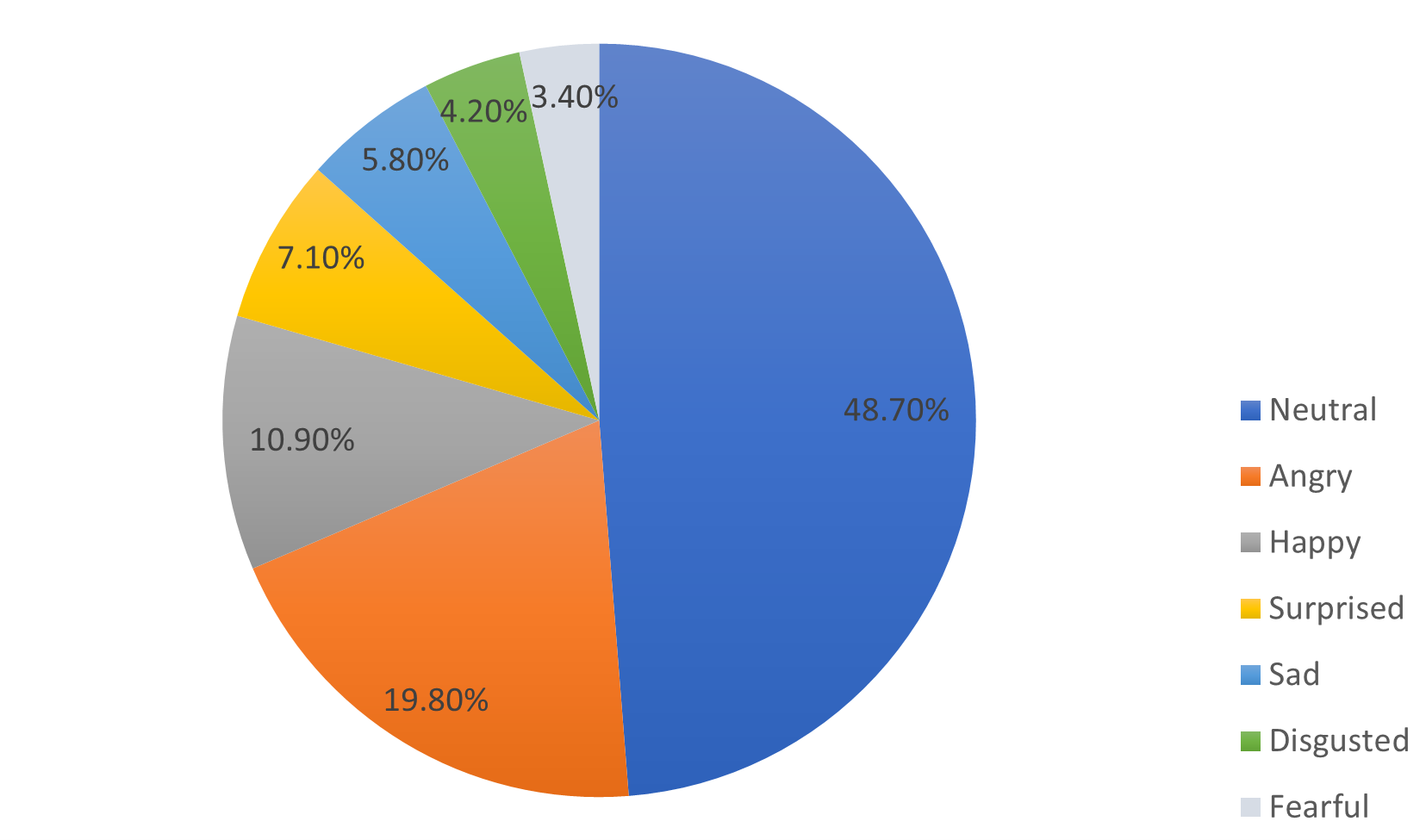}
        \caption{Audio modality data.}
        \label{fig:subfig1-1}
    \end{subfigure}
    \centering
    \begin{subfigure}{0.37\linewidth}
        \centering
        \includegraphics[width=1\linewidth]{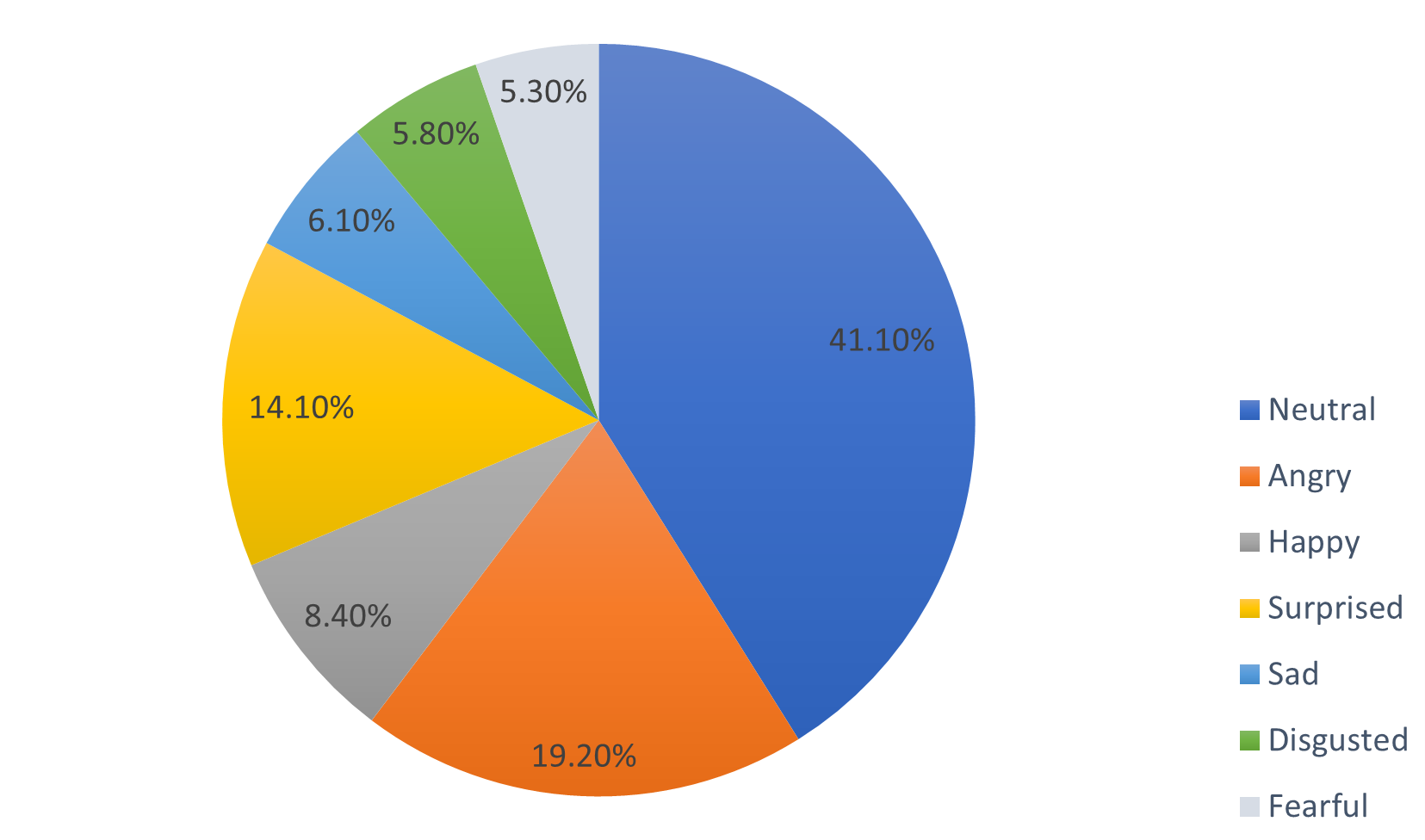}
        \caption{Text modality data.}
        \label{fig:subfig1-2}
    \end{subfigure}\\
    \centering
    \begin{subfigure}{0.37\linewidth}
        \centering
        \includegraphics[width=1\linewidth]{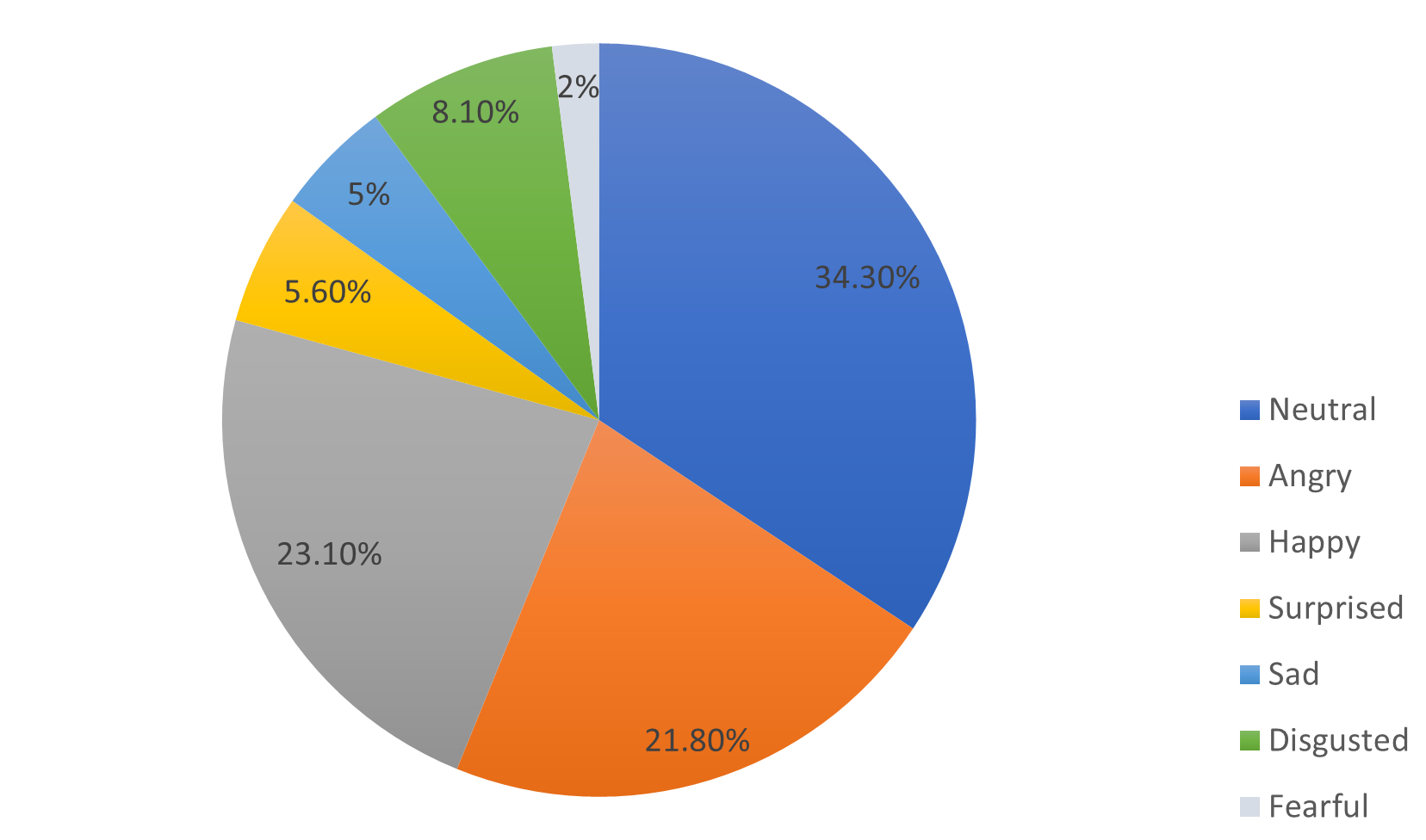}
        \caption{Video modality data.}
        \label{fig:subfig1-3}
    \end{subfigure}
        \begin{subfigure}{0.37\linewidth}	
        \centering
        \includegraphics[width=1\linewidth]{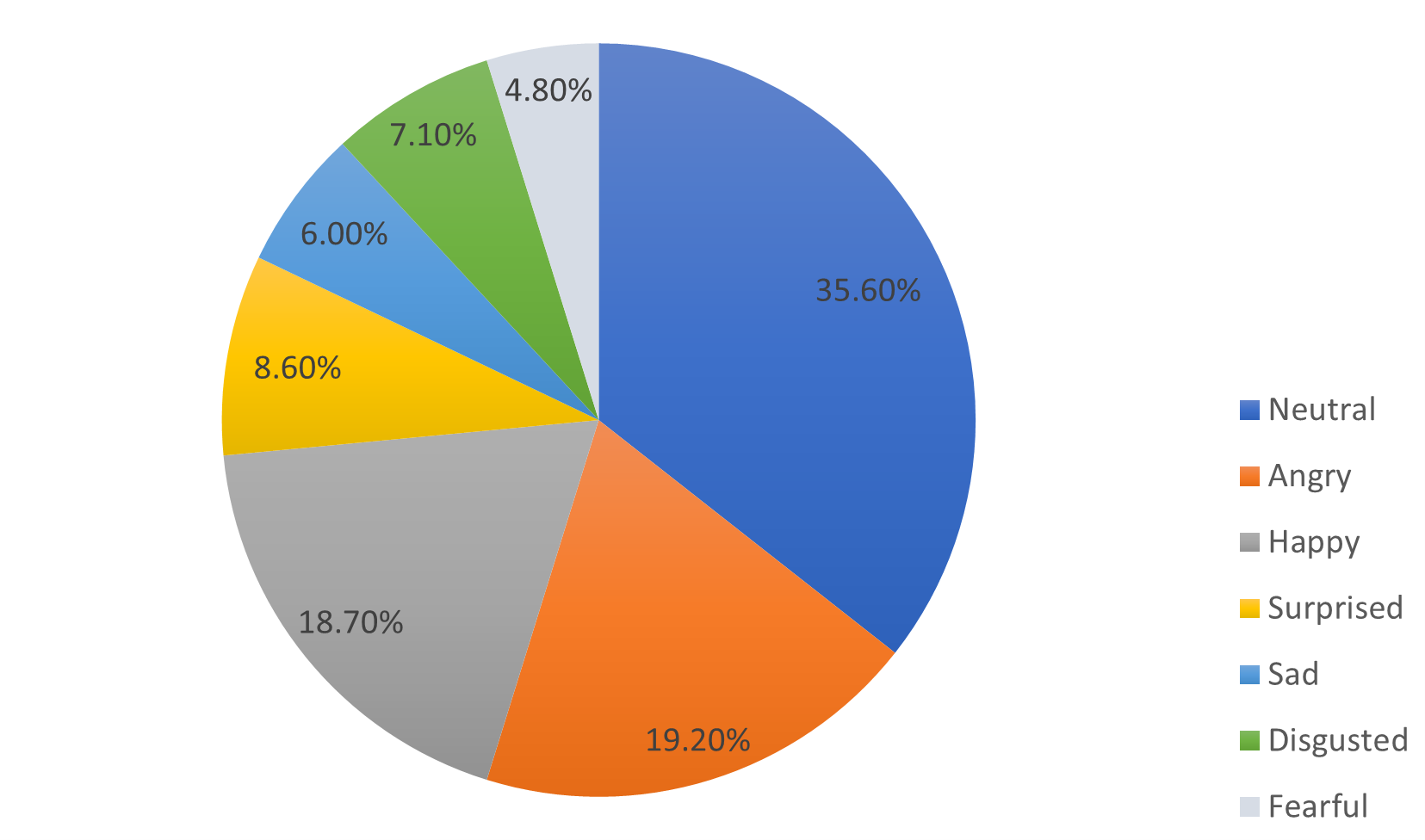}
        \caption{Multi-modality data.}
        \label{fig:subfig1-4}
    \end{subfigure}
    \caption{Data distribution across different modalities.}
    \label{statistics}
 \vspace{-6pt}
\end{figure}

In the audio modality, emotional categories exhibit a highly imbalanced distribution. As shown in Fig ~\ref{fig:subfig1-1}, neutral instances dominate, with 9,378 samples accounting for 48.7\% of the total. Angry emotions are the second most common, appearing in 3,820 instances (19.8\%). Happy emotions are found in 2,105 instances (10.9\%), followed by surprised (1,363 instances, 7.1\%), sad (1,110 instances, 5.8\%), disgusted (818 instances, 4.2\%), and fearful (656 instances, 3.4\%). Overall, neutral and angry emotions constitute the majority of the dataset.

As shown in Fig ~\ref{fig:subfig1-2}, in the text modality, 7,903 instances are labeled as neutral, accounting for 41.1\% of the total. Angry emotions comprise 3,698 instances (19.2\%), while surprised emotions account for 2,712 instances (14.1\%). The happy, sad, disgusted, and fearful categories contain 1,620 (8.4\%), 1,183 (6.1\%), 1,111 (5.8\%), and 1,023 (5.3\%) instances, respectively. Overall, the distribution of emotional categories is imbalanced, with neutral and angry emotions representing a relatively large portion of the dataset.

For the video modality, we separately analyzed the emotion recognition results for videos without and with audio, as shown in Fig \ref{fig:subfig1-3} and \ref{fig:subfig1-4}, respectively. In the video modality without audio, neutral emotions constitute the largest proportion, accounting for 6,608 instances (34.3\%). Happy and angry emotions also represent significant portions, with 4,451 (23.1\%) and 4,190 (21.8\%) instances, respectively. The remaining emotional categories include disgusted (1,563 instances, 8.1\%), surprised (1,084 instances, 5.6\%), sad (964 instances, 5.0\%), and fearful (390 instances, 2.0\%). Overall, the emotional distribution is notably concentrated in the neutral, happy, and angry categories.

For the multi-modal video modality incorporating audio, neutral emotions constitute the largest category, with 6,853 instances, representing 35.6\% of the dataset. Subsequently, angry and happy emotions are the next most frequent, occurring in 3,698 instances (19.2\%) and 3,592 instances (18.7\%), respectively. The remaining emotional categories are surprised (1,661 instances, 8.6\%), disgusted (1,371 instances, 7.1\%), sad (1,153 instances, 6.0\%), and fearful (922 instances, 4.8\%). In summary, the distribution is characterized by a significant proportion of neutral, angry, and happy emotions.

Concerning the continuous emotion labels within the multi-modal data, our analysis demonstrates a marked dominance of negative sentiments. Weakly negative and negative emotions together comprise 78.4\% of the total (49.1\% and 29.3\%, respectively). Neutral sentiments account for 11.5\%, while positive sentiments (weakly positive at 8.2\% and positive at 2.0\%) represent a smaller portion at 10.2\%. This distribution underscores a notable bias toward negative emotional expressions in the dataset, which may be attributed to the inherent characteristics of the video content or the nature of user responses.

Fleiss' Kappa is a statistical measure used to assess the agreement among multiple raters or evaluators when categorizing items into distinct categories. Fleiss' Kappa can be applied to situations with more than two raters. The formula for Fleiss' Kappa is as follows:
\[
\kappa = \frac{P_o - P_e}{1 - P_e}
\]
where \(P_o\) represents the observed proportion of agreement, and \(P_e\) represents the expected proportion of agreement under random conditions. 

The Kappa value ranges from -1 to 1. A value close to 1 indicates a high degree of consistency, a value close to 0 indicates consistency comparable to random consistency, and a negative value means consistency lower than the random level. Generally, Fleiss' Kappa is interpreted as follows: values between 0.00 and 0.20 indicate poor agreement, 0.21 to 0.40 signify fair agreement, 0.41 to 0.60 suggest moderate agreement, 0.61 to 0.80 reflect substantial agreement, and values between 0.81 and 1.00 indicate almost perfect agreement.

In our dataset, the Fleiss' Kappa value for the audio data is 0.79, for the text data is 0.66, for the video data without audio is 0.73, and for the video data with audio is 0.78. These Fleiss' Kappa values indicate good agreement across all modalities. 

\section{Experiments}
\label{sec:experiments}
In this section, we evaluate our dataset across a variety of tasks, including unimodal emotion recognition, multimodal emotion recognition, multimodal emotion analysis, and emotional speaking style captioning. We build our experimental pipeline upon the MerBench~\cite{lian2024merbench}, which provides a standardized setup for benchmarking multimodal models.

Specifically, in the continuous setting, we focus on a binary classification task that distinguishes between positive and negative emotions, where samples with scores below 0 are labeled as negative, and those above 0 as positive. For the first three tasks, accuracy (ACC) is used as the primary evaluation metric, while for the speaker emotion-style captioning task, we adopt BLEU$_4$, ROUGE$_L$, METEOR, SPIDEr, FENSE, BERTScore and CLAPScore for evaluation. To facilitate reproducibility, we document all experimental settings in Appendix~\ref{app:hyperparameters}, including hyperparameter tuning strategies, optimizer selection, and the values of all key training parameters.

\newcommand{\fontsizess}{\fontsize{9}{11}\selectfont} % 自定义表格字体大小
\begin{table*}
    \fontsizess
    \centering
\begin{tabularx}{\textwidth}{lccccc}
\toprule
% \multicolumn{6}{c}{\textbf{Speech Modality}}\\ \hline
\textbf{Speech Model} & \textbf{Speech(Four)} & \textbf{Multimodal(Four)}&\textbf{Speech(All)} & \textbf{Multimodal(All)} &\textbf{Mean}\\ \midrule
Whisper-Base  & 71.03 & 60.44 & 56.61 & 48.47 & 59.14\\
Whisper-Large  & 75.45 & 61.90 & 60.34 & 49.56 & 61.81\\
WavLM-Base  &72.50&	62.96&	59.72&	53.14	&62.08\\
Wav2vec 2.0-Base  &77.31& 63.85 &62.16&	50.96&	63.57\\
Wav2vec 2.0-Large &76.22&	64.68&	63.14&	51.06&	63.78\\
WavLM-Large &76.67&	64.48&	61.90&	53.91	&64.24 \\
Hubert-Large &\textbf{82.88} &\textbf{73.69} &66.15&	61.12&	70.96\\
Hubert-Base  &81.09& 73.09 &\textbf{68.64}&	\textbf{62.52}&	\textbf{71.34}\\ \midrule
% \multicolumn{6}{c}{\textbf{Text Modality}}\\ \hline
\textbf{Text Model} & \textbf{Text(Four)} & \textbf{Multimodal(Four)}&\textbf{Text(All)} & \textbf{Multimodal(All)} &\textbf{Mean}\\ \midrule
Vicuna-7B &55.24 &46.26 &45.57 &43.91 &47.75\\ 
LERT-Base &59.68 &51.36 &46.09 &38.26 &48.85\\
DeBERTa-Large &57.46 &49.11 &44.89 &44.79 &49.06 \\
BERT-Base &57.66 &50.83 &46.50 &44.69 &49.92\\
Sentence-BERT &56.52 &52.15 &46.45 &45.05 &50.04 \\
BLOOM-7B &60.87 &50.56 &47.38 &43.23 &50.51\\
ChatGLM2-6B& \textbf{60.95}&55.47&46.19&41.16&50.94\\
RoBERTa-Large &59.48 &53.88 &46.86 &44.27 &51.12 \\
RoBERTa-Base &60.15 &50.96 &48.11 &\textbf{45.52} &51.19\\
Baichuan-7B &60.08 &\textbf{56.39} &\textbf{48.21} &41.84 & \textbf{51.63}\\
\midrule
% \multicolumn{6}{c}{\textbf{Visual Modality}}\\ \hline
\textbf{Visual Model} & \textbf{Visual(Four)} & \textbf{Multimodal(Four)}&\textbf{Visual(All)} & \textbf{Multimodal(All)} &\textbf{Mean}\\ \midrule
Data2vec-Base &35.72 &29.69 &40.44 &32.92 &34.69 \\
VideoMAE-Base &54.18 &47.51 &54.33 &46.29 &50.58 \\
EVA-02-Base &69.87 &54.27 &58.84 &38.88 &55.47 \\
VideoMAE-Large &62.36 &64.74 &55.68 &50.54 &58.33 \\
CLIP-Base &71.38 &63.95 &59.51 &49.09 &60.98 \\
Dinov2-Large &70.60 &68.99 &60.96 &\textbf{54.59} &63.79 \\
Dinov2-Giant &73.42 &69.58 &62.73 &53.76 &64.87 \\
CLIP-Large &\textbf{77.81} &\textbf{73.96} &\textbf{64.75} &54.17 & \textbf{67.67} \\
\bottomrule
\vspace{-25pt}
\end{tabularx}
\caption{We report unimodal results for the EmotionTalk dataset. Four means that only four emotion labels are used: happy, angry, sad, and neutral. All means that all of the emotion labels are used.}
\label{uni}
\end{table*}

\subsection{Unimodal Emotion Recognition}
This section reports the emotion recognition performance of different feature extractors on the corresponding modalities, as shown in Table~\ref{uni}.

\textbf{Feature Extractor:} To assess the performance of our dataset, we employ a comprehensive suite of pre-trained baseline models across different modalities. Specifically, for the speech modality, we utilize Wav2Vec 2.0~\cite{baevski2020wav2vec}, HuBERT~\cite{hsu2021hubert}, WavLM~\cite{chen2022wavlm}, and Whisper~\cite{radford2023robust}. For the text modality, our selection includes Vicuna-7B~\cite{chiang2023vicuna}, LERT~\cite{cui2022lert}, DeBERTa~\cite{he2020deberta}, BERT~\cite{devlin2019bert}, Sentence-BERT~\cite{reimers2019sentence}, BLOOM-7B~\cite{workshop2022bloom}, RoBERTa~\cite{liu2019roberta}, ChatGLM2~\cite{du2021glm} and Baichuan-7B~\cite{yang2023baichuan}. For the visual modality, we adopt Data2Vec~\cite{baevski2022data2vec}, VideoMAE~\cite{tong2022videomae}, EVA-02~\cite{fang2024eva}, CLIP~\cite{radford2021learning}, and DINOv2~\cite{oquab2023dinov2}. 

Based on the comparative performance of these encoders shown in Table~\ref{uni}, we aim to provide guidance for modality-specific feature selection in downstream emotion recognition tasks. Given that the EmotionTalk dataset provides independent unimodal annotations, we conduct two experimental settings to investigate the capability of unimodal representations in emotion recognition. In the first setting, we utilize ground-truth unimodal labels to evaluate each model’s ability to perform unimodal emotion classification. In the second setting, we adopt multimodal labels instead, to assess whether a single modality alone can reliably infer the speaker’s actual emotional state.

Several key findings emerge from these experiments. First, for the same unimodal classification task, models consistently achieve better performance when trained and evaluated with unimodal labels than with multimodal labels. This suggests that models are effective at capturing the modality-specific emotional cues. However, these results do not necessarily reflect the speaker's actual emotional state, as unimodal annotations may be biased or incomplete. This indicates that unimodal information remains a valuable signal for emotion recognition, though it is inherently limited in expressiveness and scope. Consequently, relying solely on unimodal representations is insufficient for accurately capturing complex emotional states, reinforcing the importance of multimodal fusion in emotion understanding.

\newcommand{\fontsizessss}{\fontsize{9}{12}\selectfont} % 自定义表格字体大小
\begin{table*}
\centering
\fontsizessss
\begin{tabular}{cccccc}
\toprule
\textbf{Features} & \textbf{Algorithms} &  \textbf{Fusion} &\textbf{Multimodal(Four)}& \textbf{Multimodal(All)} &\textbf{Mean}\\ \midrule
\multirow{9}{*}{\begin{tabular}[c]{@{}l@{}}Hubert-Base\\ \\Baichuan-7B\\ \\CLIP-Large \end{tabular}}&MCTN	&Frame-level &65.34 &	47.80	&56.57	\\
&MFM 	&Frame-level &75.94	&59.51&67.73 \\
&GMFN	&Frame-level &76.87& 63.66&	70.27	\\
&MMIN	&Uttrance-level &78.93	&64.54	&71.74\\
&MISA	&Uttrance-level &80.58&	66.77&	73.68\\
&TFN	&Uttrance-level &80.12	&68.27	&74.20\\
&MulT	&Frame-level& 82.17	&66.67	&74.42\\
&MFN 	&Frame-level &80.38	&\textbf{69.31}&74.85\\
&Attention&Uttrance-level &	82.11	&68.17	&75.14\\
&LMF 	&Uttrance-level& \textbf{81.31}& 69.10 &\textbf{75.21}\\
\bottomrule
\vspace{-15pt}
\end{tabular}
\caption{We report multimodal results for the EmotionTalk dataset. Four means that only four emotion labels are used: happy, angry, sad, and neutral. All means that all of the emotion labels are used.}
\label{fusion}
\end{table*}

\newcommand{\fontsizesss}{\fontsize{7.5}{11}\selectfont} % 自定义表格字体大小

\begin{table}
\centering
\fontsizesss
\begin{tabular}{cccccccc}
\toprule
\multicolumn{8}{c}{Multimodal}\\ \midrule
\textbf{\# Top} &Text &Speech&Visual& \textbf{Discrete(Four)}& \textbf{Discrete(All)} & \textbf{Continuous} &\textbf{Mean}\\ \midrule
Top 1 &Baichuan-7B&Hubert-Base&CLIP-Large& 81.31	&69.10	&\textbf{93.35}	&81.25 \\
Top 2 &RoBERTa-Base&Hubert-Large &Dinov2-Giant & \textbf{83.23}	&\textbf{69.21}	&93.16	&\textbf{81.87}      \\
Top 3 &RoBERTa-Large&WavLM-Large&Dinov2-Large&  78.13	&65.01	&93.10	&78.75 \\ 
Top 4 &ChatGLM2-6B&W2v 2.0-Large&CLIP-Base&73.82& 63.50 &92.26& 76.53\\   
\bottomrule
\vspace{-20pt}
\end{tabular}
\caption{“Top4” indicates that we select the top 4 models for each modality (their ranking is based on the results in Table~\ref{uni}). We utilize the LMF for multimodal fusion.}
\label{multi}
\end{table}

\subsection{Multimodal Emotion Recognition / Sentiment Analysis}
Table~\ref{fusion} presents the performance of various multimodal fusion algorithms on the EmotionTalk dataset using the optimal encoder from each modality—HuBERT-Base (speech), Baichuan-7B (text), and CLIP-Large (visual). The fusion methods are categorized into frame-level (e.g., MFN~\cite{zadeh2018memory}, GMFN~\cite{zadeh2018multimodal}, MCTN~\cite{pham2019found}, MFM~\cite{tsai2018learning}, and MulT~\cite{tsai2019multimodal}) and utterance-level (e.g., TFN~\cite{zadeh2017tensor}, LMF~\cite{liu2018efficient}, MISA~\cite{hazarika2020misa}, MMIM~\cite{han2021improving}, and the Attention mechanism~\cite{vaswani2017attention}) strategies, enabling a comparative analysis of their effectiveness in multimodal emotion recognition.

Several important observations can be drawn. First, utterance-level fusion methods generally outperform frame-level approaches in both the four-class and full-class emotion classification settings. For instance, LMF achieves the highest score in the Multimodal(Four) setting (83.04\%) and also yields the best average performance (75.53\%), indicating that aligning features at the utterance level better captures the holistic emotional state. Similarly, attention-based fusion also performs competitively, with an average score of 75.14\%, suggesting the advantage of adaptive weighting across modalities. Moreover, due to the limited scale of emotion datasets, complex fusion algorithms are prone to overfitting. In contrast, simple yet effective fusion strategies often achieve relatively better performance.

Table~\ref{multi} reports the multimodal emotion recognition results using the top four models from each modality, selected based on unimodal performance in Table~\ref{uni}. All combinations adopt the LMF algorithm for fusion. Among the configurations, the combination of RoBERTa-Base (text), HuBERT-Large (speech), and Dinov2-Giant (visual) achieves the best overall performance, with the highest score in the Discrete (Four) setting (83.23\%) and the highest average (81.87\%). Notably, different model combinations yield comparable performance on the continuous labels, while their results on discrete tasks vary considerably, underscoring the impact of feature selection. These findings confirm that even under the same fusion strategy, the choice of multimodal features can significantly affect the overall performance of multimodal fusion.

\newcommand{\tablefontsize}{\fontsize{7.3}{11}\selectfont} 
\begin{table*}[t]
\tablefontsize
\centering
\begin{tabular}{ccccccccc}
\toprule
&\textbf{Decoder} & \textbf{BLEU$_4$} & \textbf{ROUGE$_L$} & \textbf{METEOR} & \textbf{SPIDEr} & \textbf{FENSE} & \textbf{BERTScore}& \textbf{CLAPScore}\\ 
\midrule
\multirow{3}{*}{Speaker} 
&Transformer-based &0.011 & 0.397 & 0.204 & 0.229 & 0.842 & 0.974 & 0.860\\
&GPT-2 &\textbf{0.020} & \textbf{0.430} & \textbf{0.212} & 0.256 & 0.765 & 0.976 & \textbf{0.899}\\
&Qwen-2 &0.009 & 0.414 & 0.205 & \textbf{0.258} & \textbf{0.846} & \textbf{0.977} & 0.878\\\midrule
\multirow{3}{*}{Style} 
&Transformer-based &0.065 & 0.517 & 0.313 & 0.339 & 0.512 & 0.985 & 0.895 \\
&GPT-2 &0.075 & 0.510 & 0.298 & 0.350 & \textbf{0.611} & 0.987 & 0.850\\
&Qwen-2 &\textbf{0.127} & \textbf{0.564} & \textbf{0.339} & \textbf{0.482} & 0.523 & \textbf{0.988} & \textbf{0.912}
\\\midrule
\multirow{3}{*}{Emotion} 
&Transformer-based &0.032 & 0.366 & 0.191 & 0.276 & 0.932 & 0.973 & 0.843\\
&GPT-2 &0.014 & \textbf{0.399} & 0.147 & 0.235 & 0.903 & 0.972 & 0.818\\
&Qwen-2 & \textbf{0.058} & 0.361 & \textbf{0.199} & \textbf{0.353} & \textbf{0.942} & \textbf{0.975} & \textbf{0.853}
\\\midrule
\multirow{3}{*}{Overall} 
&Transformer-based &0.018 & 0.469 & 0.233 & \textbf{0.230} & \textbf{0.921} & 0.980 & 0.878\\
&GPT-2 &0.015 & 0.462 & 0.214 & 0.227 & 0.890 & 0.980 & 0.849\\
&Qwen-2 & \textbf{0.033}& \textbf{0.535}& \textbf{0.268}& 0.121& 0.562& \textbf{0.984}& \textbf{0.885}\\
\bottomrule
\vspace{-20pt}
\end{tabular}
\caption{Automatic captioning results. All methods use Hubert as the speech encoder.}
\label{tab:description}
\end{table*}

\subsection{Emotional Speaker Style Captioning}
Table~\ref{tab:description} presents a comprehensive comparison of three decoder architectures—Transformer-based, GPT-2, and Qwen-2—across multiple captioning dimensions (Speaker, Style, Emotion, and Overall), evaluated using a suite of standard automatic metrics. Qwen-2 outperforms other models across all four tasks, demonstrating its effectiveness in producing captions that preserve both emotional nuance and stylistic diversity. The strong BERTScore suggests that its generation aligns closely with human references at the semantic level, beyond surface-level lexical similarity. Although Qwen-2 achieved the best overall performance, GPT-2 performed notably well on the speaker-focused task, obtaining the highest ROUGE$_L$ (0.430) and CLAPScore (0.899). In contrast, the Transformer-based decoder showed weaker overall results but maintained basic structural coherence and content coverage, as reflected in its ROUGE$_L$ and SPIDEr scores. Overall, these results highlight Qwen-2’s robustness in capturing fine-grained stylistic and emotional cues, crucial for emotional speaking style captioning.

\section{Conclusion}
In this paper, we construct an interactive Chinese multimodal sentiment dataset, EmotionTalk, filling the gap in Chinese multimodal emotion research that lacks high-quality recorded data and emotional speaking style captions dataset. The dataset includes 23.6 hours of multimodal data recorded by 19 professional actors from the drama department, ensuring the authenticity and naturalness of emotional expressions. EmotionTalk is the first interactive multimodal emotion dataset in Chinese with fine-grained labels and emotional speaking style caption annotations, making it a valuable resource for affective computing community. In addition, we conduct experiments on unimodal emotion recognition, multimodal emotion recognition and emotional speaking style caption tasks to validate the quality of the dataset. The dataset is freely available to the academic community, aiming to promote further development in the fields of affective computing and human-computer interaction, and to provide valuable resource support for related research.

\bibliographystyle{plainnat}
\bibliography{neurips_2025}

\newpage
\appendix
\section{Datasheets for datasets}
\subsection{Dataset Snapshots}
The dataset comprises 744 dialogues, encompassing a total of 19,250 utterances for each unimodal modality—text, audio, and video. The audio data span approximately 23.6 hours, with an average duration of 4.4 seconds per utterance.

\begin{figure}[htbp]
    \centering
    \begin{subfigure}{1.0\linewidth}
        \centering
        \includegraphics[width=1\linewidth]{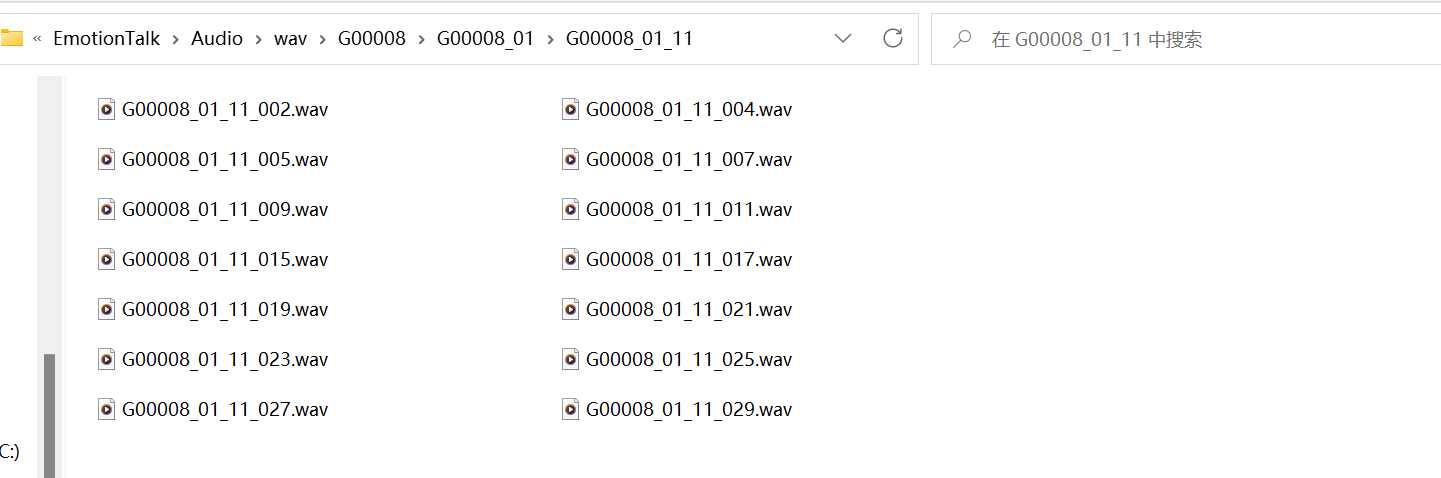}
        \caption{Examples of audio file samples.}
        \label{fig:subfig1}
    \end{subfigure}
    \centering
    \begin{subfigure}{1.0\linewidth}
        \centering
        \includegraphics[width=1\linewidth]{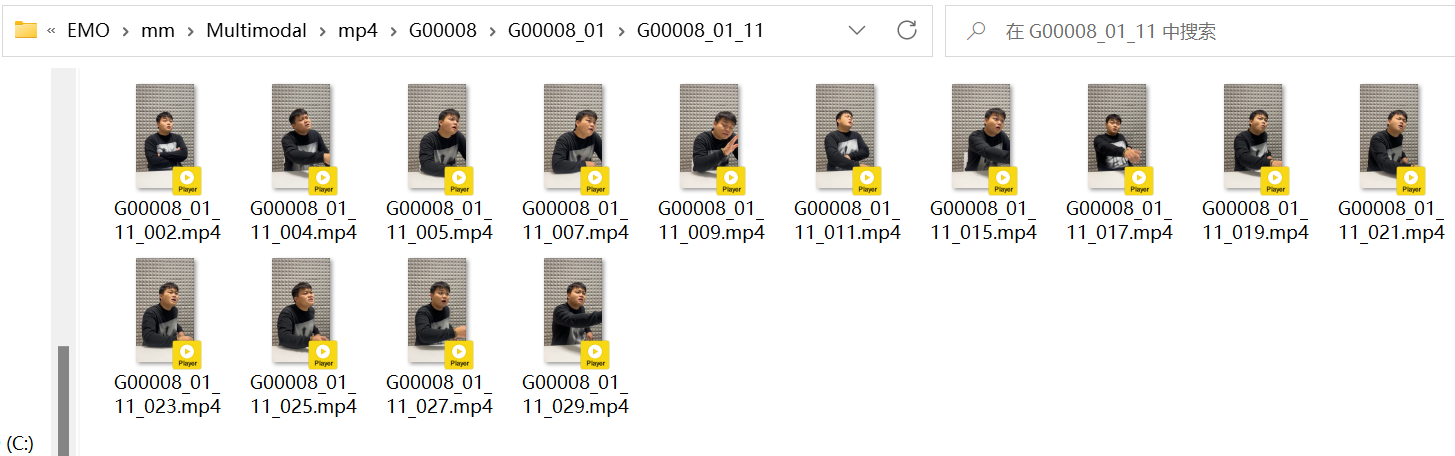}
        \caption{Examples of video file samples.}
        \label{fig:subfig2}
    \end{subfigure}\\
    \caption{Snapshots of audio and video samples in the EmotionTalk dataset. All files are named following a consistent and structured format.}
    \label{snapshot}
\end{figure}

Each utterance is stored as an individual JSON file following a unique naming convention in the format: "<group\_No>\_<session\_No>\_<Speaker\_id>\_<Utt\_No>.json". Corresponding audio and video files are named identically, with the extensions ".wav" and ".mp4" respectively: "<group\_No>\_<session\_No>\_<Speaker\_id>\_<Utt\_No>.wav" and "<group\_No>\_<session\_No>\_<Speaker\_id>\_<Utt\_No>.mp4". The samples of audio and video files in EmotionTalk are shown in Fig ~\ref{snapshot}.

\subsection{Data Format}
Each utterance in the EmotionTalk dataset is associated with a corresponding ".jsonl" file, which contains detailed sample-level annotations. These annotations include not only the basic information such as the emotion label, speaker identity, and transcript, but also rich metadata that describes the expressive characteristics of the utterance. The detailed annotation fields are listed in Table~\ref{tab:sample_annotations}. % An example of a ".jsonl" file for a single utterance is shown in Table~\ref{tab:sample_jsonl}.

\subsection{Data Distribution}
\label{app:distribution}
In this study, to make full use of the data and ensure both effective model training and fair evaluation, the dataset is divided into training, validation, and test sets in a approximate ratio of 8:1:1. Specifically, 80\% of the data is used for training the model to learn effective feature representations, 10\% is allocated for validation to assist in model selection and prevent overfitting during training, and the remaining 10\% is reserved as the test set to evaluate the model’s generalization performance. When splitting the dataset, we make effort to ensure that the data distribution of each category remains consistent across the training, validation, and test sets. A detailed information of the the distribution across different subsets is presented in the Table ~\ref{tab:dataset_split}.

\begin{table}[htbp]
\renewcommand{\arraystretch}{1.2}  % 调整行间距
\centering
\begin{threeparttable}
\caption{Description of Sample-Level Annotations}
\begin{tabular}{ll}
\toprule
\textbf{Name} & \textbf{Description}\\
\midrule
emotion & Emotion label.\tnote{1}  \\
Confidence\_degree   & Annotator's self-rated confidence in the emotion label. \\
Continuous\_label  & 5-dimensional sentiment labels.\tnote{2} \\
speaker\_id  & Unique speaker identifier. \\
emotion\_result  & Final aggregated emotion label.\tnote{3} \\
Continuous label\_result & Final averaged sentiment labels aggregated from five annotators. \\
content  & Transcript of the utterance. \\
startTime  & Utterance start time in the session. \\
endTime  & Utterance end time in the session. \\
duration  & Total duration of the utterance. \\
emo\_cap  & Caption describing the type and intensity of the expressed emotion. \\
spe\_cap  & Caption describing the speaker's voice quality. \\
style\_cap  & Caption describing speaking style. \\
caption\_1~--~caption\_5  &  Emotional speaking style caption. \\
file\_path  & Relative path to the audio file. \\
\bottomrule
\end{tabular}
\begin{tablenotes}
\footnotesize
\item[1] The emotion categories include: happiness, surprise, sadness, disgust, anger, fear, and neutral.
\item[2] The 5-dimensional sentiment labels include: -2 (strongly negative), -1 (weakly negative), 0 (neutral), 1 (weakly positive), or 2 (strongly positive).
\item[3] The computation method is detailed in Section~3.2 Annotation.
\end{tablenotes}
\label{tab:sample_annotations}
\end{threeparttable}
\end{table}
\begin{table}[htbp]
\centering
\caption{Statistics of the data distribution across the training, validation, and test sets.}
\begin{tabular}{lcccccccc}
\toprule
\textbf{} & \textbf{Angry} & \textbf{Disgusted} & \textbf{Fearful} & \textbf{Happy} & \textbf{Neutral} & \textbf{Sad} & \textbf{Surprised} & \textbf{Total} \\
\midrule
Train & 2950 & 1142 & 672 & 2986 & 5377 & 919 & 1367 & 15413 \\
Validation   & 409  & 95   & 125 & 360  & 675  & 111 & 133  & 1908  \\
Test  & 339  & 134  & 125 & 246  & 801  & 123 & 161  & 1929  \\
\midrule
Total & 3698 & 1371 & 922 & 3592 & 6853 & 1153 & 1661 & 19250 \\
\bottomrule
\end{tabular}
\label{tab:dataset_split}
\end{table}
\section{Feature Extraction}
\subsection{Models}
\label{app:models}
To comprehensively evaluate the proposed dataset, we conduct extensive experiments on three tasks: unimodal emotion recognition, multimodal emotion recognition / sentiment analysis and emotional speaker style captioning. For the unimodal and multimodal emotion recognition tasks, we employ a range of state-of-the-art models as feature extractors to obtain representations from each modality. Then, we select several high-quality features as the foundation for multimodal fusion. For the emotional speaker style captioning task, we utilize three types of decoders, including transformer, GPT-2 and Qwen-2, to assess the quality and utility of the dataset. The details of the models are provided in Table ~\ref{tab:models}.

\subsection{Hyperparameters and computing resources}
\label{app:hyperparameters}
We provide open access to both the data and the code used in our experiments. The full experimental code is available at \url{https://github.com/NKU-HLT/EmotionTalk}. Key training hyperparameters for different models are summarized in Table~\ref{hyperparams1}, Table~\ref{fusion_hyperparams_reordered}, and Table~\ref{caption_hyper}. All models are trained using the AdamW optimizer.

The experiments based on the Qwen-2 decoder are conducted on an NVIDIA A800 GPU, while all other experiments are performed using an NVIDIA GeForce RTX 3090 GPU.

\newcommand{\ttablefontsize}{\fontsize{8.3}{11}\selectfont} 
\begin{table}[htbp]
\ttablefontsize
\centering
\caption{An overview of the models employed across different tasks.}
\begin{tabular}{lll}
% \begin{tabular}{p{5cm}p{8cm}}
\toprule
\textbf{Speech Model} & \textbf{Link} & \textbf{License} \\
\midrule
Whisper-Base~\cite{radford2023robust} & huggingface.co/openai/whisper-base & Apache License 2.0\\
Whisper-Large~\cite{radford2023robust} & huggingface.co/openai/whisper-large-v2 & Apache License 2.0\\
WavLM-Base~\cite{chen2022wavlm} & huggingface.co/microsoft/wavlm-base & CC BY-SA 3.0\\
Wav2vec 2.0-Base~\cite{baevski2020wav2vec} & huggingface.co/TencentGameMate/chinese-wav2vec2-base & MIT License \\
Wav2vec 2.0-Large~\cite{baevski2020wav2vec} & huggingface.co/TencentGameMate/chinese-wav2vec2-large & MIT License\\
WavLM-Large~\cite{chen2022wavlm} & huggingface.co/microsoft/wavlm-large & CC BY-SA 3.0\\
Hubert-Large~\cite{hsu2021hubert} & huggingface.co/TencentGameMate/chinese-hubert-large & MIT License\\
Hubert-Base~\cite{hsu2021hubert} & huggingface.co/TencentGameMate/chinese-hubert-base & MIT License\\
\midrule
\textbf{Text Model} & \textbf{Link} & \textbf{License}\\
\midrule
Vicuna-7B~\cite{chiang2023vicuna} & huggingface.co/CarperAI/stable-vicuna-13b-delta & CC BY-NC-SA 4.0\\
LERT-Base~\cite{cui2022lert} & huggingface.co/hfl/chinese-lert-base & Apache License 2.0\\
DeBERTa-Large~\cite{he2020deberta} & huggingface.co/microsoft/deberta-v3-large & MIT License\\
BERT-Base~\cite{devlin2019bert} & huggingface.co/google-bert/bert-base-chinese & Apache License 2.0 \\
Sentence-BERT~\cite{reimers2019sentence} & huggingface.co/sentence-transformers/paraphrase-multilingual & Apache License 2.0\\
& -mpnet-base-v2 \\
BLOOM-7B~\cite{workshop2022bloom} & huggingface.co/bigscience/bloom-7b1 & BigScience Responsible \\ & & AI License 1.0\\
ChatGLM2-6B~\cite{du2021glm} & huggingface.co/THUDM/chatglm2-6b & Apache License 2.0\\
RoBERTa-Large~\cite{liu2019roberta} & huggingface.co/hfl/chinese-roberta-wwm-ext-large 
& Apache License 2.0\\
RoBERTa-Base~\cite{liu2019roberta} & huggingface.co/hfl/chinese-roberta-wwm-ext & Apache License 2.0\\
Baichuan-7B~\cite{yang2023baichuan} & huggingface.co/baichuan-inc/Baichuan-7B \\
\midrule
\textbf{Visual Model} & \textbf{Link} & \textbf{License}\\
\midrule
Data2vec-Base~\cite{baevski2022data2vec} & huggingface.co/facebook/data2vec-vision-base & Apache License 2.0\\
VideoMAE-Base~\cite{tong2022videomae} & huggingface.co/MCG-NJU/videomae-base & CC BY-NC 4.0 \\
EVA-02-Base~\cite{fang2024eva} & https://huggingface.co/timm/eva02\_base\_patch14\_224.mim\_in22k & MIT License\\
VideoMAE-Large~\cite{tong2022videomae} & huggingface.co/MCG-NJU/videomae-large & CC BY-NC 4.0\\
CLIP-Base~\cite{radford2021learning} & huggingface.co/openai/clip-vit-base-patch32 & Apache License 2.0\\
Dinov2-Large~\cite{oquab2023dinov2} & huggingface.co/facebook/dinov2-large & Apache License 2.0\\
Dinov2-Giant~\cite{oquab2023dinov2} & huggingface.co/facebook/dinov2-giant & Apache License 2.0\\
CLIP-Large~\cite{radford2021learning} & huggingface.co/openai/clip-vit-large-patch14 & Apache License 2.0\\
\midrule
\textbf{Captioning Model} & \textbf{Link} & \textbf{License}\\
\midrule
Transformer-based~\cite{lewis2019bart} & huggingface.co/fnlp/bart-base-chinese & Apache License 2.0\\
GPT-2~\cite{lagler2013gpt2} & huggingface.co/uer/gpt2-chinese-cluecorpussmall & Apache License 2.0\\
Qwen-2~\cite{yang2407qwen2} & huggingface.co/Qwen/Qwen2-7B & Apache License 2.0\\
\bottomrule
\end{tabular}
\label{tab:models}
\end{table}

\begin{table}[b!]
\centering
\small
\begin{tabular}{lcc}
\toprule
\textbf{Hyperparameter} & \textbf{Four (Unimodal/Multimodal)} & \textbf{All (Unimodal/Multimodal)} \\
\midrule
Learning Rate & 1e-3 & 1e-5 \\
L2 Regularization Weight & 1e-5 & 1e-5 \\
Batch Size & 32 & 32 \\
Epochs & 100 & 100 \\
\bottomrule
\end{tabular}
\caption{Training hyperparameters used for the unimodal and multimodal models in Table 2 on the EmotionTalk dataset. "Four" refers to using four emotion labels (happy, angry, sad, neutral), while "All" refers to using the full label set. }
\label{hyperparams1}
\end{table}
\begin{table*}[h]
\centering
\small
\begin{tabular}{lcccccc}
\toprule
\textbf{Model} & \textbf{Hidden Dim} & \textbf{Dropout} & \textbf{Learning Rate} & \textbf{Grad Clip} \\
\midrule
MCTN      & 64 – 256       & 0.0 – 0.3     & 1e-3     & 0.6 – 1.0  \\
MFM       & 128 / 256      & 0.0 – 0.7     & 1e-3     & -1.0  \\
GMFN      & 128 / 256      & 0.0 – 0.7     & 1e-3     & -1.0  \\
MMIN      & 64 – 256       & 0.0 – 0.3     & 1e-3     & 0.6 – 1.0 \\
MISA      & 64 – 256       & 0.2 – 0.5     & 1e-4     & -1.0 – 1.0  \\
TFN       & 64 / 128       & 0.2 – 0.5     & 1e-3     & -1.0  \\
MulT      & 64 – 256       & 0.0 – 0.3     & 1e-3     & 0.6 – 1.0  \\
MFN       & 128 / 256      & 0.0 – 0.7     & 1e-3     & -1.0  \\
Attention & 64 – 256       & 0.2 – 0.5     & 1e-5     & -1.0 \\
LMF       & 32 – 256       & 0.2 – 0.5     & 1e-5     & -1.0 \\
\bottomrule
\end{tabular}
\caption{Key training hyperparameters used for each multimodal model in Table 3 on the EmotionTalk dataset.}
\label{fusion_hyperparams_reordered}
\end{table*}
\begin{table}[h!]
\centering
\small
\begin{tabular}{lccccc}
\toprule
\textbf{Decoder} & \textbf{Batch Size} & \textbf{Epochs} & \textbf{Learning Rate} & \textbf{Weight Decay} & \textbf{Warmup} \\
\midrule
Transformer-based & 8  & 15 & 1.7e-05 & 3.0e-04 & 0 \\
GPT-2             & 8  & 15 & 1.7e-05 & 3.0e-04 & 0 \\
Qwen-2            & 4  & 6  & 1e-4 & 0.0 & 1,000 \\
\bottomrule
\end{tabular}
\caption{Training hyperparameters for each decoder in Table 5.}
\label{caption_hyper}
\end{table}

\begin{figure}[b!]
    \centering
    \begin{subfigure}{0.77\linewidth}
        \centering
        \includegraphics[width=1\linewidth]{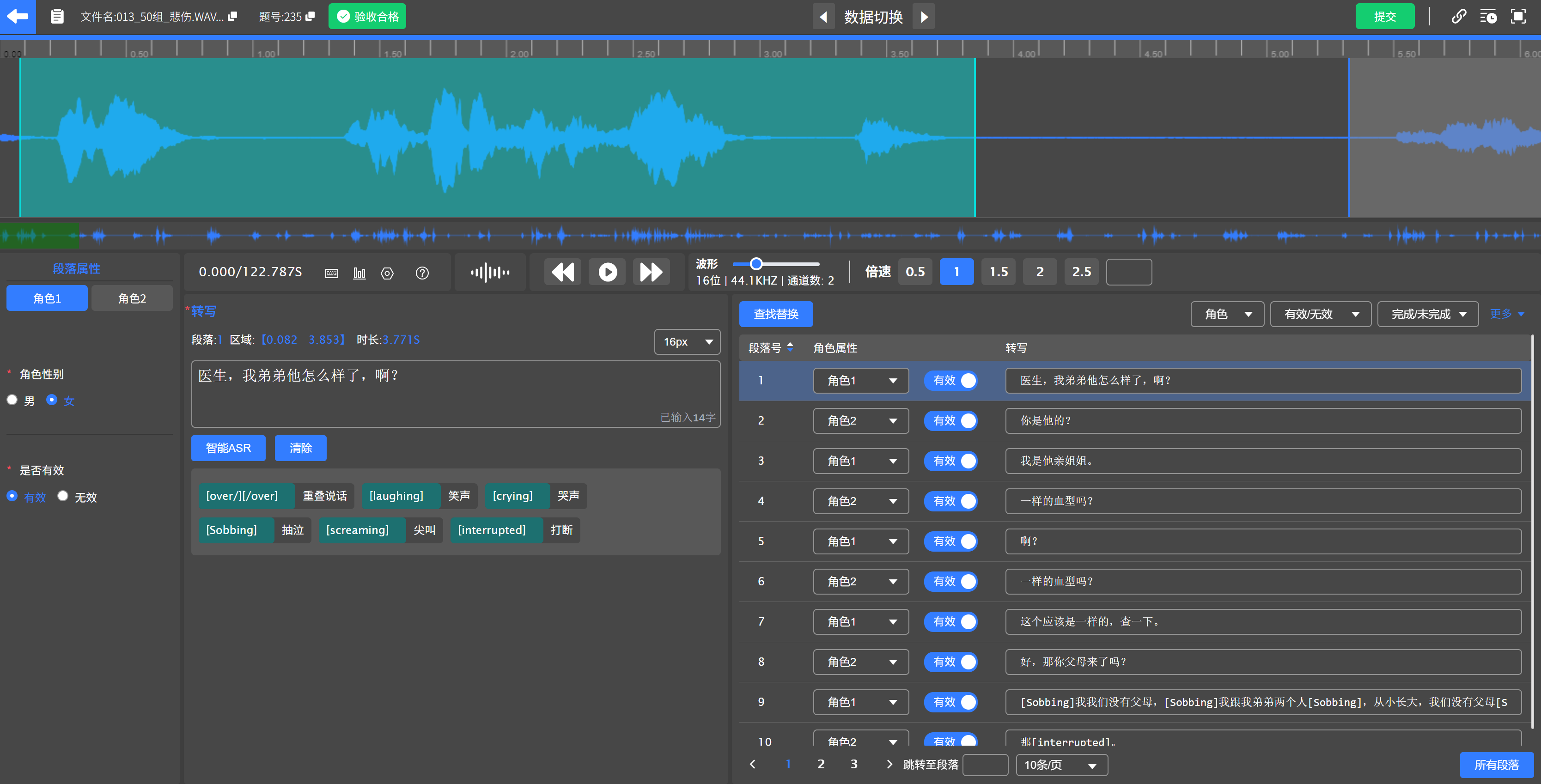}
        \caption{Annotation platform of the speech emotion recognition.}
        \label{fig:subfig1}
    \end{subfigure}
    \centering
    \begin{subfigure}{0.77\linewidth}
        \centering
        \includegraphics[width=1\linewidth]{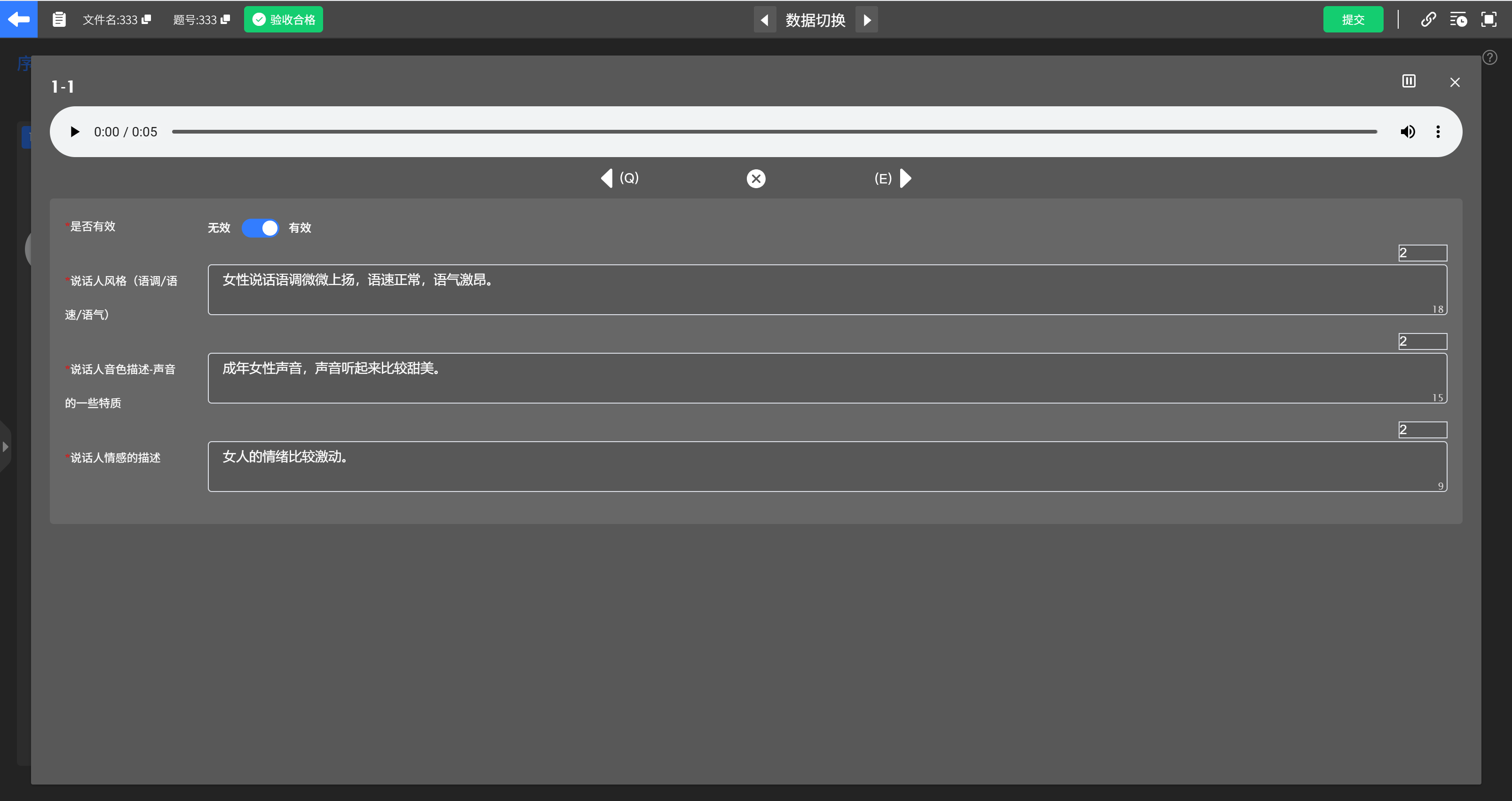}
        \caption{Annotation platform of the emotional speaking style captioning.}
        \label{fig:subfig2}
    \end{subfigure}\\
    \caption{Overview of the annotation platform interface.}
    \label{platform}
\end{figure}
\section{Annotation Website}

To improve the efficiency of the annotation process, we conduct data annotation and quality assessment on a data platform. As shown in the Fig ~\ref{platform}, this platform supports the annotation of various tasks such as speech emotion recognition and emotional speaking style captioning.

\section{Extra Experiment Results and Analysis}
\renewcommand{\tablefontsize}{\fontsize{7.3}{11}\selectfont} 
\begin{table}[htbp]
\tablefontsize
\centering
\caption{Accuracy (\%) comparison of different models across modalities (Speech, Text, and Speech+Text) on the Emotion Prediction in Conversation (EPC) task.}
\label{tab:epc-results}
\begin{tabular}{lcccc}
\toprule
\textbf{Modality} & \textbf{Model} & \textbf{ACC} \\
\midrule
\multirow{4}{*}{Speech}
  & BiGRU & 62.58 \\
  & AVEF  & 60.81 \\
  & DEP   & 65.40 \\
  & EAMT  & \textbf{65.85} \\
\midrule
\multirow{4}{*}{Text}
  & BiGRU & 44.19 \\
  & AVEF  & \textbf{49.92} \\
  & DEP   & 49.53 \\
  & EAMT  & 48.68 \\
\midrule
\multirow{4}{*}{Speech+Text}
  & BiGRU & \textbf{65.70} \\
  & AVEF  & 64.47 \\
  & DEP   & 65.25 \\
  & EAMT  & 65.47 \\
\bottomrule
\end{tabular}
\end{table}

\begin{table}[htbp]
\tablefontsize
\centering
\caption{Accuracy (\%) comparison of different models across modalities (Speech and Text) on the Emotion Recognition in Conversation (ERC) task.}
\label{tab:erc-results}
\begin{tabular}{lcc}
\toprule
\textbf{Modality} & \textbf{Model} & \textbf{ACC} \\
\midrule
\multirow{4}{*}{Speech}
  & CMN           & 66.37 \\
  & ICON          & 65.31 \\
  & DialogueRNN   & 66.34 \\
  & DialoguGCN    & \textbf{67.51} \\
\midrule
\multirow{4}{*}{Text}
  & CMN           & 46.67 \\
  & ICON          & 47.38 \\
  & DialogueRNN   & 49.63 \\
  & DialoguGCN    & \textbf{49.75} \\
\bottomrule
\end{tabular}
\end{table}

In the task of Emotion Prediction in Conversation (EPC), we observe clear performance differences across modalities and models~\cite{shahriar2019audio, shi2020dimensional, shi2023emotion}, as shown in Table~\ref{tab:epc-results}. Speech-based models consistently outperform text-based ones, highlighting the importance of vocal information in anticipating upcoming emotional states. Among the speech-only models, EAMT~\cite{shi2023emotion} achieves the highest accuracy at 65.85\%, marginally surpassing DEP~\cite{shi2020dimensional} (65.40\%) and BiGRU (62.58\%), indicating the benefit of explicitly modeling multi-role contextual information. Text-only models perform significantly worse, with BiGRU scoring only 44.19\%, and EAMT, despite its contextual modeling, reaching just 48.68\%. Interestingly, multimodal fusion (speech+text) does not yield substantial improvements over speech alone. The BiGRU model achieves 65.70\% when both modalities are used, only slightly better than its speech-only counterpart. Similarly, DEP and EAMT see marginal, suggesting that the textual signal may offer limited complementary information for EPC.

In Emotion Recognition in Conversation (ERC), a similar modality gap is evident, with speech-based models outperforming their text-based versions by a large margin~\cite{hazarika2018conversational, yeh2019interaction, majumder2019dialoguernn, ghosal2019dialoguegcn}, as shown in Table~\ref{tab:erc-results}. DialoguGCN~\cite{ghosal2019dialoguegcn} achieves the best performance among all models with 67.51\% accuracy, leveraging its graph-based structure to effectively capture speaker interactions and contextual flow. DialogueRNN~\cite{majumder2019dialoguernn} and CMN~\cite{hazarika2018conversational} also perform competitively in the speech modality (66.34\% and 66.37\%, respectively), while ICON~\cite{yeh2019interaction} slightly lags behind at 65.31\%. In contrast, their text-based counterparts yield significantly lower results—DialoguGCN at 49.75\% and DialogueRNN at 49.63\%—underscoring the limitations of relying solely on lexical information for emotion recognition. Notably, DialoguGCN consistently outperforms the other models across both modalities, suggesting its architectural advantage in handling complex conversational dynamics. 

All models are trained using the Adam optimizer with an initial learning rate of 1e-4 and a weight decay of 1e-5. The batch size is set to 16, and models are trained for a maximum of 30 epochs (except for DialogueRNN, which is trained for 100 epochs). A StepLR scheduler is employed to decay the learning rate by a factor of 0.1 every 10 epochs in ERC.

\section{Ethics Statement and License}
\label{app:ethics}
This study is conducted in accordance with rigorous ethical guidelines to ensure the protection of participants' rights and well-being. All recordings take place in a quiet indoor environment, where professional actors engage in natural, emotionally diverse, and logically coherent dialogues based on predefined emotional themes and content outlines. The annotation cost for each data sample is 0.2 RMB.

To preserve participant privacy, all data are anonymized by removing personal identifiers and replacing them with coded labels. The dataset is released under the CC BY-NC 4.0 license, which prohibits commercial use and supports ethical research practices. Data are securely stored, and access is restricted to authorized researchers for academic purposes only.

In conclusion, this study demonstrates a strong commitment to ethical standards, encompassing informed consent, the protection of personal privacy, appropriate compensation, and the responsible dissemination of data, thereby safeguarding participant rights and supporting ethical scientific advancement.

\section{Accessibility}
\label{app:accessibility}
The EmotionTalk dataset will be released soon.

\section{Impact}
\label{app:impact}
\subsection{Positive Impact}
This paper introduces \textbf{EmotionTalk}, first interactive multimodal emotion dataset in Chinese, with fine-grained labels and emotional speaking style caption annotations. EmotionTalk addressing the lack of high-quality Chinese datasets in the field of multimodal emotion research. We conduct a series of experiments on unimodal, multimodal, and emotional speaking style captioning tasks to assess the quality of the dataset. The EmotionTalk dataset has been publicly released with the goal of advancing research in affective computing.

\subsection{Negative Impact}
Although the EmotionTalk dataset comprises 23.6 hours of recordings, its scale remains subject to further enhancement. Additionally, there exists a problem of the imbalanced category distribution in our dataset. Overall, the neutral emotion category contains the largest number of samples, while the fearful emotion category has the fewest. While this distribution is more consistent with real-world emotional occurrences, it can still constrain the model’s ability to accurately identify certain emotion categories.

\section{Limitations}
\label{app:limitations}
Although the dataset includes 23.6 hours of multimodal data and holds significant value for MMER, its scale is still relatively small compared to the large datasets used in fields such as speech synthesis and speech recognition. This limits the applicability of the dataset in large-scale data processing and model training. Additionally, the dataset is recorded by 19 professional actors from different regions. While this ensures the naturalness and diversity of emotional expression, the relatively small number of participants and the uneven geographical distribution may affect the representativeness of the data. In the future, expanding the number of participants and incorporating samples from a broader range of regions and cultural backgrounds will help further enrich the dataset's diversity, enhance its adaptability to different emotional expressions and cultural contexts, and thereby improve the effectiveness and applicability of sentiment analysis models in a wider range of real-world scenarios.

\end{document}